\documentclass[fleqn,usenatbib]{mnras}

\usepackage[dvipsnames]{xcolor}
\usepackage{multirow}
\usepackage{newtxtext,newtxmath}
\usepackage{color,soul}
\newcommand{\referee}[1]{{#1}}

\usepackage[T1]{fontenc}

\DeclareRobustCommand{\VAN}[3]{#2}
\let\VANthebibliography\thebibliography
\def\thebibliography{\DeclareRobustCommand{\VAN}[3]{##3}\VANthebibliography}

\usepackage{graphicx}	
\usepackage{subcaption}
\usepackage{float}
\usepackage{amsmath}	

\title[Orbital Behaviour of Hildas I: Stability]{On the Long-Term Orbital Behaviour of Hilda Asteroids - I. Global Stability Patterns and Catalogue}

\author[C. F. Chavez et al.]{
Cristian F. Chavez,$^{1,2,3}$\thanks{E-mail: astroengineer.com@gmail.com}
Jonathan Horner,$^{1}$
Patryk Sofia Lykawka$^{4}$
Timothy R. Holt$^{1}$
Brad Carter$^{1}$
\\
$^{1}$Centre for Astrophysics, University of Southern Queensland, West St, Darling Heights, Toowoomba, QLD 4350, Australia\\
$^{2}$Centro de Investigaci\'on y Desarrollo en Ciencias Aeroespaciales, Academia Polit\'ecnica Aeron\'autica, Fuerza A\'erea de Chile, JMC 11087, Santiago, Chile\\
$^{3}$Centro Espacial Nacional, Fuerza A\'erea de Chile, Av. Pedro Aguirre Cerda 5500, Santiago, Chile\\
$^{4}$Kindai University, Shinkamikosaka 228-3, Higashiosaka, Osaka 577-0813, Japan\\
}

\date{Accepted XXX. Received YYY; in original form ZZZ}

\pubyear{\the\year{}}

\begin{document}
\label{firstpage}
\pagerange{\pageref{firstpage}--\pageref{lastpage}}
\maketitle

\begin{abstract}
The Hildas are objects trapped in 3:2 mean-motion resonance (MMR) with Jupiter, and are a fascinating window into the Solar system's formation and evolution. We present the results of detailed dynamical integrations of 2,699 numbered asteroids in the Hilda region, following 243 clones of each Hilda for 1 Gyr under the influence of the giant planets and the Sun; 
54.7\% of our sample exhibit extreme dynamical stability, with all clones of 1,475 Hildas remaining trapped in the 3:2 MMR for the full 1 Gyr of our integrations. A further 408 Hildas (15.1\% of the sample) see more than 50\% of their clones surviving within the MMR at the end of the simulations. Of the remaining 30.2\% of the population, 327 objects (12.1\%) lose 50\% of their clones at some point between the 100 Myr and 1 Gyr point, leading to dynamical half-lives measured in hundreds of millions of years -- a result compatible with those objects being members of a once larger population of Hildas trapped since the Solar system's youth. The final 489 objects (18.12\%) prove highly unstable -- and may well be interlopers, captured to the Hilda population in the relatively recent past. 
Overall, our results suggest that the Hilda population was larger in the past, while its remarkably high stability over Gyr timescales is consistent with a primordial origin.
 
\end{abstract}




\section{Introduction}

The Solar system is the one planetary system that we can study in exquisite detail, and, as such, the knowledge and insights gained from it provide the critical underpinning of our understanding of how planetary systems form and evolve over time. Over the past few decades, our understanding of the Solar system's evolution has increased dramatically, with old ideas of a gentle and relatively static formation history \citep[e.g.][]{Edge1949,Lissauer93} being replaced by the modern dynamic and more violent narrative, with planets falling victim to giant collisions \citep[e.g.][]{Earth_1,Merc_1,Can01,AH8,Earth_3,Merc_2,Can18,Liu19}, and migrating over large distances from their formation locations to where we observe them today \citep[e.g.][]{nice1,gomes,migrate2,lykawka,NT2,NT3,migrate3,migrate4}\footnote{For a detailed overview of our current understanding of the Solar system's formation and evolution, and particularly the role of the Solar system's small bodies in building that understanding, we direct the interested reader to the excellent review articles by \citet{migrate4}, \citet{SSRev}, and \citet{GladRev}, and references therein.}.

The study of the Solar system's small body populations has proved key to this journey of scientific discovery. In particular, the Solar system's populations of resonant small bodies have provided vital new insights into the migration of the Solar system's gas and ice giant planets. The peculiar orbit of the dwarf planet Pluto, and the orbits of the other `Plutinos', trapped in 2:3 mean-motion resonance with the gas giant Neptune, is best explained by the idea that planet has migrated outwards since its formation, traveling more than a billion kilometres, trapping and transporting the Plutinos as it did so \citep[e.g.][]{Malhotra93,Malhotra95,CJ02,VM19}. The migration of Neptune is further supported by studies of the Neptune Trojans, trapped in 1:1 mean-motion resonance with the ice giant, which suggest that the known population was captured and excited to their current orbits during Neptune's outward journey \citep[e.g.][]{lykawka,NT2,NT3,Parker15,GN16}. 

Closer to the Sun, the Jovian Trojan population has been a particular focus of scrutiny when it comes to disentangling the chaotic migration history of the gas giants Jupiter and Saturn \citep[e.g.][]{morbi05,LH10,nesv13,migrate4,nesv18b,pirani19}. It is now generally accepted that the Jovian Trojans are a captured population, acquired by Jupiter towards the end of its migration, likely aided by interactions between Jupiter and Saturn as the two planets crossed through 5:2 mean-motion resonance. As a result, the Jovian Trojan population likely contains objects with a wide variety of compositions that formed at a diverse range of heliocentric distances -- from material that originated in the Asteroid belt and inner Solar system to objects that formed beyond the orbit of Neptune \citep[e.g.][]{morbi05,nesv13,Dones15,Emery15}\referee{, an idea that is strongly supported by observational studies \citep[e.g.][]{Wong15,Bely24,Ver25}}.

The distribution of asteroids in the Asteroid belt (and particularly those located close to mean-motion and secular resonances) has also provided a wealth of information that has been used to study the migration of the giant planets \citep[e.g.][]{MinMal09,Morbi10,secularsaturn,Walsh12,Deie16,IZ16}, providing clues to both the scale and chaoticity of the final stages of planetary migration. As with the Jovian Trojans, there is growing evidence that material from the outer Solar system (that formed in the trans-Neptunian region) was emplaced in the Asteroid belt during planetary migration \citep[e.g.][]{LevCont,vol16,LI23,NesTric24}. In particular, observations of the dwarf planet (1) Ceres have revealed a volatile-rich world, abundant in materials that infer a formation location in the outer Solar system. In particular, the presence of ammoniated phyllosilicates requires the presence of ammonia on the dwarf planet in its youth, suggesting an origin in the trans-Neptunian region \citep[e.g.][]{Ceres15,Russell16}.


In this context, the population of Solar system small bodies known as the Hildas are of particular interest. Trapped in mean-motion resonance with Jupiter, several thousand potential Hilda asteroids have been identified in the 150 years since the discovery of the first and archetypal object, (153) Hilda, in 1875 \citep{1876-Hilda}. The Hildas are located beyond the outer edge of the broad torus of the classical main asteroid belt, with instantaneous semi-major axes that range between $\sim3.7$ and $\sim4.2$ au. 

The effect of the Jovian 3:2 mean-motion resonance provides a mechanism by which the Hildas can maintain stable orbits on timescales comparable to the age of the Solar system \citep[e.g.][]{Ferraz-Melo-GI1998, broz, chavez1}, despite moving on orbits that can bring them relatively close to Jupiter's orbit. When stable Hildas approach aphelion, and come closest to Jupiter's orbit, the giant planet is far away; approximately sixty degrees ahead or behind the location of the Hilda in its orbit, or on the opposite side of the Sun. In other words, when the Hildas are near aphelion, they approach the locations of the L$_3$, L$_4$, and L$_5$ Jovian Lagrange points. 

When those Hildas and Jupiter are closest to one another, the Hildas are typically near perihelion, maximising their distance from the giant planet. When the locations of the Hildas are plotted in the Solar system, this results in a distinctive orbital distribution pattern, with the Hildas appearing to occupy, and to move along, the sides of an equilateral triangle that rotates along with the orbital motion of Jupiter. The long-term evolution of the Hildas within the 3:2 resonance can act to maintain this stable architecture -- with perturbations from Jupiter gradually changing the asteroid's orbit in such a way that close encounters between the two are prevented, and ensuring that the Hilda's aphelia continue to align with the Jovian L$_3$, L$_4$, and L$_5$ Lagrange points.

Compared to the Jovian Trojans, the origins of the Hilda population remain relatively poorly studied. \citet{Franklin-Hildas-capturados2004} investigated the eccentricity distribution of the Hildas, arguing that the best way to reproduce the observed distribution was to consider the Hildas as a population of objects captured during the inward (sunward) migration of Jupiter. As a result, they argued that the Hildas, and other resonant asteroids in the outer main belt, could serve as excellent probes of Jupiter's migration. This formation mechanism is supported by observational studies that compare the Hilda and Jovian Trojan populations \citep[e.g.][]{Marsset14,wong17,spectra}, finding that the two populations display broadly similar observational properties that are in keeping with a shared origin. 

More recently, \citet{Vokrouhlicky-2025-Hildas-mag-dist} performed an extensive study of the Hildas, allowing them to compare the absolute magnitude distribution of the Hildas to that of the Jovian Trojans. They find that the slope of the Hilda's magnitude distribution is markedly shallower than that of the Jovian Trojans of the same size, which suggests that the true population of Hildas is an order of magnitude smaller than that of the Trojans, despite the fact that the two populations contain roughly the same number of large objects. This is challenging for models that suggest a capture origin for those populations, as such studies suggest that the captured populations of Trojans and Hildas should be roughly comparable in size \citep{vol16}, though they note that differences in the collisional histories of the two populations may play a role in explaining the observed differences, as was suggested by \citet{wong17}. 

\cite{broz} performed a detailed study of asteroids trapped in first-order resonance with Jupiter. In the course of that study, they identified 1197 members of the Hilda group, finding clear evidence of two potential collisional families in that population. They found that (1911) Schubart was accompanied by a collisional family with an estimated age of 1.7$\pm$0.7 Gyr, and identified a looser collisional family associated with (153) Hilda, whose age they estimated at $\gtrsim$ 4 Gyr.

Building on that work, \citet{Broz-Hildas-LHB} investigated (153) Hilda's collisional family, finding once again that it was formed approximately 4 Gyr ago. In addition, they calculated the collisional rates in the current Hilda population, with their results suggesting that the probability of a disruptive impact on an object the size of (153) Hilda would be low across the age of the Solar system. Such a finding is in keeping with numerical and statistical studies that suggest that the Hildas were captured to their current orbits during the latter stages of planet formation, and that they could share a common origin with the Jupiter Trojans \citep[e.g.][]{trojan-hilda-sim-origin,Gaspar,terai,SlizBalogh-2023_Jup_Organizer}. Such periods of chaotic capture would clearly lead to increased collision rates throughout the inner Solar system, increasing the likelihood of disruptive impacts in the Hilda region.

More recently, \cite{Vino19} identified two small collisional families associated with the Hildas (1212) Francette and (2483) Guinevere, in a study that identified a total of 141 such families across the breadth of the Asteroid belt and Jovian resonant populations. \citet{Vokrouhlicky-2025-Hildas-mag-dist} confirmed the existence of those two families, identifying 151 members of the (1212) Francette family, and 54 members of the (2483) Guinevere family. In addition, they found two further collisional families amongst the Hildas -- the (1345) Potomac family, with more than 500 identified members, and the (269345) 2008 TG$_{106}$ family, with 17 members.

It is clear that the Hildas offer a fascinating window into the formation and evolution of the Solar system. If they are a captured population, then it seems likely that, just like the Jovian and Neptunian Trojans, they will contain objects that formed at different locations in the Solar system, and that have experienced markedly different histories. Indeed, \citet{chavez1} used data obtained using the Faint Object infraRed CAmera for the SOFIA Telescope to obtain observations of the two of the largest Hildas, (1162) Larissa and (1911) Schubart, at mid-infrared wavelengths. That data was used to characterise the physical properties and thermal inertia of the two objects. The study revealed two highly disparate objects, with very different albedoes and thermal properties -- though whether their differences are innate, or the result of the collision process that birthed the (1911) Schubart collisional family, remains open to debate.

In this work, we present the results of a detailed suite of $n$-body simulations investigating the dynamical stability of 2,699 numbered asteroids whose orbits meet the criteria to be considered potential members of the Hilda population. In Section~\ref{Sec:Simulations}, we describe the methodology behind our simulations, before presenting the results of our dynamical stability analysis in Section~\ref{Sec:Stability}. Finally, we discuss our results and draw our conclusions in Section~\ref{Sec:Conclusions}.

\begin{figure*}
    \centering

    \begin{minipage}[t]{0.475\textwidth}
        \centering
        \includegraphics[
            height=0.47\textheight,
            keepaspectratio
        ]{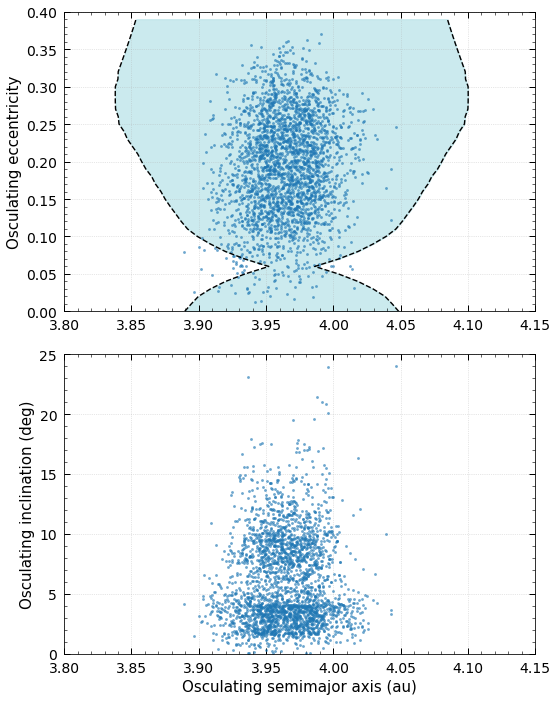}
    \end{minipage}
    \hfill
    \begin{minipage}[t]{0.475\textwidth}
        \centering
        \includegraphics[
            height=0.47\textheight,
            keepaspectratio
        ]{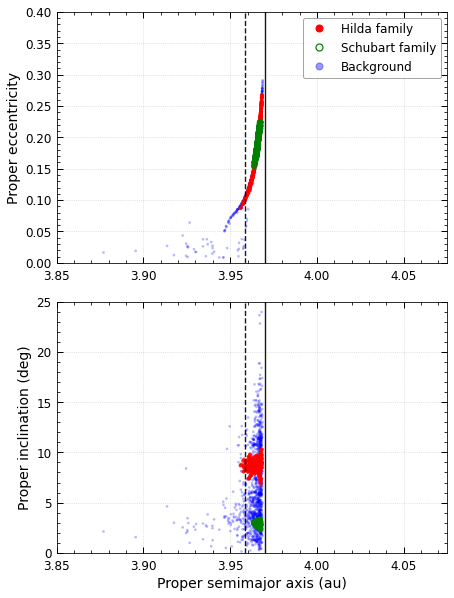}
    \end{minipage}

    \caption{Orbital distribution of the Hilda asteroids considered in this work. The left-hand panels show the initial osculating orbital elements of the complete sample of 2,699 numbered asteroids, taken from the AstDyS database and used as initial conditions for our numerical integrations, whilst the right-hand panels show the synthetic proper elements for the objects in our sample for which these quantities were available in the AstDyS catalogue (2,073 asteroids). The upper panel displays semimajor axis versus eccentricity, while the lower panel displays semimajor axis versus inclination. Members of the Hilda and Schubart collisional families are shown in red and green, respectively, in the right-hand panels, while the remaining objects are shown in blue. The solid and dashed black vertical lines in the right-hand panels indicate the nominal locations of the 3:2 mean-motion resonance with Jupiter and the 15:4 mean-motion resonance with Saturn, respectively. The shaded area in the upper left-hand panel shows the region encompassed by the 3:2 resonance with Jupiter for an orbital inclination of 5.5$^\circ$, the mean value for our sample of simulated objects, calculated using the SUPERATLAS code detailed in \citet{Gallardo2019} and \citet{Gallardo2020}.}
    \label{fig:orbital_elements}
\end{figure*}

\begin{figure*}
    \centering
    \includegraphics[width=0.49\textwidth]{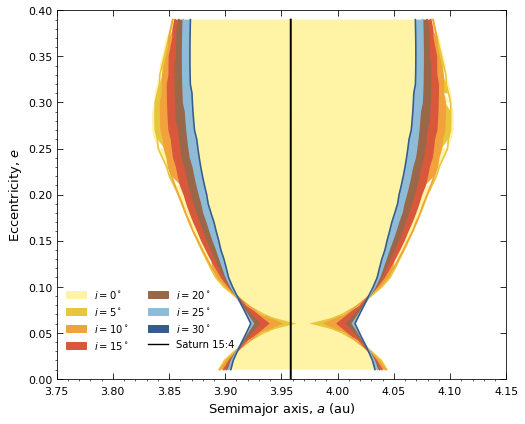}
    \hfill
    \includegraphics[width=0.49\textwidth]{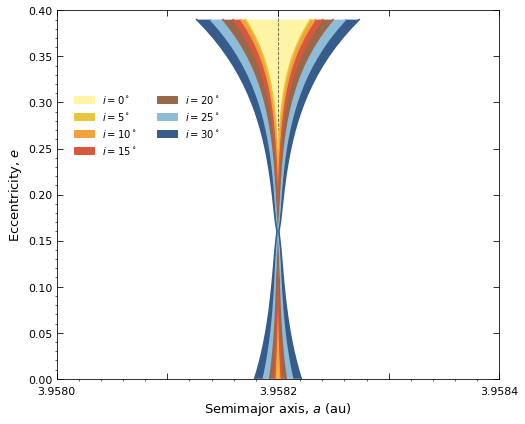}
    \caption{The left-hand panel shows the semi-analytical structure of the Jovian 3:2 and Saturnian 15:4 mean-motion resonances in the $(a,e)$ plane, calculated using the resonance model of \citet{Gallardo2020}. The variation of the stable resonance domains for fixed orbital inclinations of $i=0^\circ, 5^\circ, 10^\circ, 15^\circ, 20^\circ, 25^\circ,$ and $30^\circ$ were calculated through the adoption of a fixed reference angular configuration of $\omega=\Omega=0^\circ$, and are shown by the different coloured regions. The Saturnian 15:4 resonance is so narrow that no structure can be seen, presenting as a single vertical line at a$\sim$3.9582 au. To show the structure of that resonance, the right-hand panel presents a zoomed-in view of the structure of the 15:4 Saturnian resonance. Both resonances were calculated independently within the semi-analytical framework.}
    \label{fig:resonance_structure}
\end{figure*}

\section{SIMULATING THE NUMBERED HILDAS}
\label{Sec:Simulations}

To analyse the orbital behaviour of the Hilda population, we performed detailed $n$-body simulations of clusters of test particles based on all numbered asteroids that could potentially be Hildas\footnote{Our sample was drawn from the population of numbered asteroids on 30 May 2023, with orbital elements taken from the NASA/JPL Solar system Dynamics website, \url{https://ssd.jpl.nasa.gov/sbdb_query.cgi}}. To build our sample of integration subjects, we searched the database of numbered asteroids for all objects with current orbital semi-major axes, $a$, between 3.7 and 4.2 au, and then further restricted our sample by additionally choosing only those objects with orbital eccentricities, $e$, less than 0.3. This process yielded a sample of 2,699 potential Hildas for our study, all of which had long observational arcs and well-refined orbits.

Once we had built our sample, we obtained the orbital elements, and associated uncertainties, for each object from the AstDys proper element database \citep{AstDys}, following the example of previous studies \citep[e.g.][]{QR322,LC18,Vinogradova-2015,Id-Ast-Nesvorny-B-C-2015,non-grav-par-Wlodarczyk-2022}. For each Hilda in our sample, we used this information to create a swarm of 243 test particles, centred on the best-fit orbital solution for the object in question. We generated clones of each Hilda in a regular 5-dimensional grid in $a-e-i-\Omega-\omega$ space, with three unique values of each variable being tested -- the best fit value for that variable, and values $\pm 1\sigma$ from that value. In this way, the cloning process created a 5-cube (a five-dimensional hypercube) of test particles, with one test particle located at each location in that structure, for a total of 243 clones\footnote{Three clones in each of five-dimensions yield a total number of clones of $3^5$ = 243.} centred on the best fit orbital solution for the object in question. 

The simulations were conducted using the Hybrid integrator from the dynamics package MERCURY \citep{mercury}, and followed the orbital evolution of all test particles under the gravitational influence of the four giant planets and the Sun for a period of 1 Gyr using an integration timestep of 60 days. Test particles were followed until they either collided with one of the massive objects, fell into the central body (with a radius of 0.005 au), or were ejected from the system. Following earlier work \citep[e.g.][]{swift,Horner-2004,Anchises,QR322,chavez1}, we considered test particles to be ejected upon reaching a barycentric distance of 1000~au. All simulations were set to begin on 2000-Jan-01 00:00:00.0000, and the orbital elements and the Cartesian locations of all objects were output at intervals of 10~kyr.\footnote{\referee{The mass of the terrestrial planets was added to the central body, and those planets were not included as individual objects in our simulations, in line with previous dynamical studies of such objects. Even the most eccentric of} \referee{the Hildas considered in this study move on orbits that remain far outside the domain of the terrestrial planets, with the closest possible encounters between those Hildas and Mars having the object separated by a distance of more than 1~au, more than 125 times the size of Mars' instantaneous Hill radius at aphelion.}}

In total, our simulations followed 655,857 individual test particles integrated over 1 Gyr. In order to quantify the scale of the numerical dataset, and allow comparison to previous work, we introduce the nominal Total Particle-Years (TPY) metric, defined as the product of the total number of integrated particles and the maximum integration time-span. For the present work, this corresponds to \(6.56\times10^{14}\) TPY. For comparison, the integrations of \citep{Dahlgren-1998__Hildas_vel} covered some \(5.0\times10^{7}\) TPY, while the long-term simulations detailed by \citep{Vokrouhlicky-2025-Hildas-mag-dist} covered \(\sim4.9\times10^{13}\) TPY. These values illustrate the significant gains that have been made in computing power in recent years, and we note in this light that our suite of simulations took approximately one year of real-time to complete on the University of Southern Queensland's High-Performance Computing Cluster, Fawkes.

Figure~\ref{fig:orbital_elements} shows comparison plots of the distribution of our sample in both osculating- and proper-element space. This complementary approach helps frame the discussion developed in the following sections, as the proper elements provide a clearer representation of the underlying population structure within the Jovian 3:2 resonant region, whilst the osculating ones depict the actual orbital states used to generate the initial conditions for our numerical integrations.

In order to provide a complete dynamical context for the orbital distribution of the Hilda population, we calculated the semi-analytical structure of the Jovian 3:2 and Saturnian 15:4 mean-motion resonances using the model of \citet{Gallardo2020}. Figure~\ref{fig:resonance_structure} shows the resulting stable resonance domains in the $(a,e)$ plane for seven fixed orbital inclinations between $0^\circ$ and $30^\circ$. The 15:4 resonance with Saturn was included as it falls within the area covered by the Jovian 3:2 resonance, though is expected to be far weaker, given its higher order. The Jovian 3:2 resonance exhibits a broad and inclination-dependent structure, whilst the high-order Saturnian 15:4 resonance is extremely narrow. As a result, that resonance can only be resolved on the enlarged semi-major axis scale shown in the right-hand panel.

\section{STABILITY ANALYSIS}
\label{Sec:Stability}

As has been the case for dynamical simulations of other populations of resonant objects in the Solar system (such as the Jovian Trojans \citep[e.g.][]{LevTroj,Tsig05,Anchises,Holt20,Greenstreet2020}, the Neptunian Trojans \citep[e.g.][]{Nesv02,MarzQR322,BrassQR322,QR322,KV18,LC18,Guan12}, and the Plutinos and Twotinos exterior to the orbit of Neptune \citep[e.g.][]{Ip97,Morb97,Mun26}), our sample of Hildas displayed a wide variety of dynamical lifetimes -- from objects where all clones remained Hildas for the whole simulation period to cases where all clones were removed from the system within less than 10 Myr. Indeed, 180 of the 2699 Hildas studied in this work were sufficiently unstable that more than 50\% of their clones were removed from our simulations within the first 10~Myr of the runs.

To help us to study the behaviour of the Hilda population as a whole in this work, we define a Hilda-specific classification scheme based on the number of surviving clones after simulation intervals of 1~Gyr, 100~Myr, and 10~Myr. This scheme is intended to support a more systematic and comparative analysis of potential Hilda-type asteroids across different dynamical lifetimes, and to allow us to group together objects that display similar dynamical behaviour. In this scheme, Hildas for which no clones are lost over the 1 Gyr simulation period are placed in Group I, with subsequent groups containing objects exhibiting ever greater levels of instability. Group V, which we call the `Interlopers', gathers together those objects for which more than 50\% of all test particles are lost in the first 10 Myr of our simulations. This classification scheme, to which we will refer in our analysis hereafter, is presented in Table~\ref{tabla:classif_scheme}, which also details the number of objects that fall into each of our categories, from our total sample of 2,699 Hildas. Whilst more than half of the objects considered in this work remained resolutely dynamically stable for the full 1 Gyr of our simulations (1475 objects in Group I - the Strongly Stable Hildas), a significant fraction ($\sim$45\%) show at least some instability through the course of our simulations. We present the distribution of objects through the five categories in Table~\ref{tabla:classif_scheme}.

The inclusion of Group~V (the Interlopers) -- objects with extremely short dynamical lifetimes resulting from the rapid loss of clones -- finds precedent in analogous studies of near-Earth short-period comets. \citet{Fernandez-2015_Jup_fam_comets} showed that certain comets with unstable orbits and frequent planetary encounters can be interpreted as \textit{dynamical interlopers}, temporarily populating orbital zones to which they do not genuinely belong. By analogy, our classification aims to distinguish Hilda-like bodies that only transiently occupy the 3:2 resonance before being rapidly ejected due to strong dynamical instability. Such objects are similar to interlopers identified in other resonant small body populations -- such as the temporarily captured Neptune Trojan 2004$_{\textrm{KV}18}$ \citep{KV18}, and ties in with previous studies that showed that members of the Centaur population can be captured temporarily in the Jovian Trojan population, sometimes for surprisingly long periods \citep[e.g.][]{Karl04,centaur_trojan,Greenstreet2020,Green24}.

A sample of the catalogue containing the detailed simulation outcomes for the 2,699 numbered Hilda asteroids studied in this work is presented in Appendix~\ref{sec:appendix_tables}.

A version of the catalogue is maintained in the project GitHub repository: \url{https://github.com/cchavez-astro/Hildas_table_cchavez}, which we will update with the results of additional simulations carried out for any future work.

\begin{table}
\centering
\caption{Classification scheme used in this work. Groups are defined by the fraction of surviving clones as a function of time. Group~I objects complete the full 1~Gyr integration without the loss of any of their 243 clones, while Group~V objects lose more than 50\% within the first 10~Myr. The rightmost column lists how many of our 2,699 numbered Hildas fall in each group, and the fraction of the total sample that represents.}
\label{tabla:classif_scheme}
\footnotesize
\setlength{\tabcolsep}{5pt}
\renewcommand{\arraystretch}{1.15}
\begin{tabular*}{\columnwidth}{@{\extracolsep{\fill}} l l r@{}}
\hline
\textbf{Group} & \textbf{Surviving clones} & \textbf{Number of Hildas} \\
\hline
I. Strongly Stable Hildas           & 100\% at 1~Gyr          & 1\,475 (54.7\%) \\
II. Stable Hildas                              & 50\%--100\% at 1~Gyr    & 408 (15.1\%)    \\
III. Moderately Stable Hildas       & 50\%--100\% at 100~Myr  & 327 (12.1\%)    \\
IV. Unstable Hildas                            & 50\%--100\% at 10~Myr   & 309 (11.4\%)    \\
V. Interlopers                                 & $<\,50\%$ at 10~Myr     & 180 (6.7\%)     \\
\hline
\end{tabular*}
\end{table}

\subsection{Surviving clones}
Our simulations reveal that a significant fraction {54.7\%} of the Hildas exhibit such strong dynamical stability that 100\% of their clones survive until the end of the 1 Gyr integration period. This outcome is consistent with previous numerical and analytical studies \citep[e.g.][]{Hildas-1976, Dahlgren1993, Franklin-Hildas-capturados2004, broz, chavez1, Asano-2024}, which emphasized the long-term stability of Hilda-type orbits despite their moderately excited orbital eccentricities and inclinations. In our case, the high survival rate observed among the clones reinforces the notion that the 3:2 mean-motion resonance with Jupiter provides a dynamically protected region capable of hosting stable configurations over gigayear timescales.

Specifically, when we divide our sample according to the detailed classification scheme presented in Table~\ref{tabla:classif_scheme}, we find:

\begin{itemize}
    \item The simulations delivered 1,475 (54.7\%) \textbf{`Strongly Stable'} objects, for which no clones are removed from the simulation through collision or ejection throughout the full 1 Gyr duration of our simulations (Group I). We therefore consider these objects to be the core of the very stable population of the Hildas, demonstrating that the 3:2 mean-motion resonance with Jupiter can be highly robust against secular perturbations 
    in the present-day Solar system. Indeed, examination of the orbital elements for all surviving clones at the 1 Gyr epoch shows that all clones of all objects in this group remained in the Hilda population at the end of our simulations.
    \\
    \item The second group, the \textbf{`Stable Hildas'} are those objects that display significant dynamical stability, whilst seeing some clones escape across the 1 Gyr integration period. These 408 objects (15.1\% of our sample) are analogous to objects like 2008 LC$_{18}$ of the Neptune Trojans \citep{LC18,Guan12} -- sufficiently dynamically stable that it is reasonable to consider them to be representatives of a once larger primordial Hilda population, whilst exhibiting enough instability to suggest a very gradual attrition of the Hildas in the aeons to come.
    \\
    \item The third group, the 327 \textbf{`Moderately Stable Hildas'}, are more interesting - exhibiting sufficiently strong instability that more than half of their population decay over the 1 Gyr period of integration. The instability of these objects is similar to that exhibited by objects like Anchises, in the Jovian Trojan population \citep{Anchises}, and 2001 QR$_{322}$ of the Neptune Trojans \citep{MarzQR322,BrassQR322,QR322} -- at the more unstable end of those objects that seem likely to be primordial. Taken together with the objects in groups one and two, we can consider that $\sim$ 81.8\% of our sample proved to be sufficiently dynamically stable to be considered likely relics of the primordial Hilda population.
    \\
    \item Group~IV comprises a set of \textbf{`Unstable Hildas}' (309 members, 11.5\%), while Group~V corresponds to a population of potential \textbf{`Interlopers'} (180 members, 6.7\%) that were found crossing the Hilda region at the time the ephemerides were retrieved. Once again, there are direct analogues for such objects in other resonant populations -- such as the temporarily captured Neptune Trojan 2004 KV$_{18}$ \citep{KV18,Guan12}, Uranus' short-lived Trojan companions (83982) Crantor \citep{delafM13} and (687170) 2011 QF$_{99}$ \citep{Alex13}, the transient retrograde Jovian co-orbital (514107) BZ$_{509}$ \citep{Greenstreet2020}, and a number of objects identified as potential quasi-satellites or horseshoe companions to the terrestrial planets \citep[e.g.][]{Wie98,Christ11,deLaFMars,Greg26}.
\end{itemize}

Taken together, the results detailed above paint a picture of a dynamically diverse Hilda population, suggesting differences among its members. This finding is consistent with recent observational work indicating that the Hilda population is far from a homogeneous ensemble. \citet{Vokrouhlicky-2025-Hildas-mag-dist} analysed the orbital and absolute magnitude distribution of Hildas using decade-long Catalina Sky Survey data, unveiling a clear size-frequency structure and identifying multiple collisional families. Their results show that the size distribution of the Hilda population exhibits a steep slope for larger objects and a shallower one for smaller bodies, consistent with a primordial origin modulated by collisional and dynamical evolution. This observed stratification reinforces the validity of our proposed classification: objects belonging to stable groups (I--III) likely represent long-term residents of the 3:2 resonance, while unstable members (Groups IV--V) could include dynamically transient interlopers and fragments from more recent disruptions.\\

\begin{figure*}
    \centering
    \includegraphics[width=\textwidth]{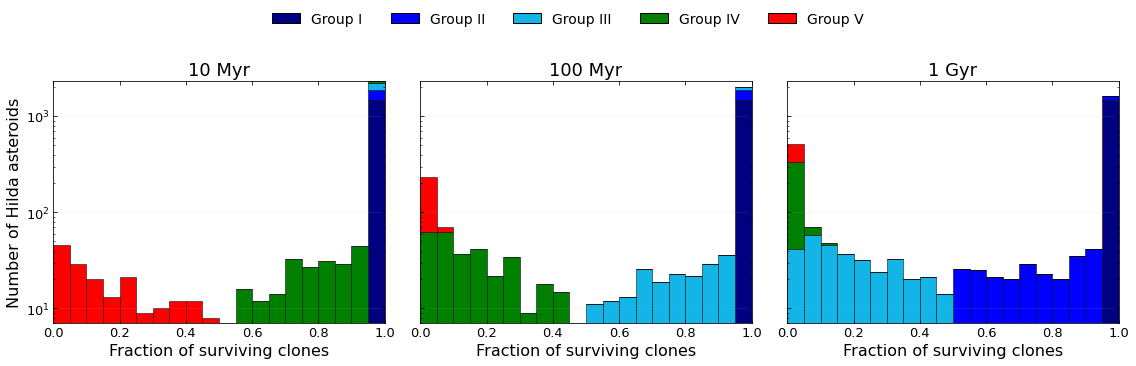}
    \caption{The distribution of test particle survival fractions for the Hilda population studied in this work after 10 Myr (left), 100 Myr (centre), and 1 Gyr (right). The survival fraction for each individual Hilda is the ratio of the number of surviving test particles to the initial 243 particles simulated for that object. Histograms are divided into bins of width 0.05 in survival fraction, corresponding to approximately 12 test particles per bin. The colours show which of the Hildas fall in the various stability categories defined in this work, ranging in stability from the least stable objects (Group V) to the most stable (Group I). The number of Hilda asteroids per bin is shown on a logarithmic scale.}
    \label{fig:histograms}
\end{figure*}

Figure~\ref{fig:histograms} presents histograms that offer an alternative view of how the different stability groups of Hildas in our study behave over time. Those histograms show how the test particle survival fractions within our sample change through the course of our integrations. For a given Hilda, at a given time, the test particle survival fraction is the ratio of the number of clones of that Hilda that remain in the simulation divided by the total number of clones that were present at the start of the simulation (243). For each Hilda in our study, the test particle survival fraction was determined after 10, 100, and 1000 Myr, and the histograms show the distribution of these survival fractions across our sample at those three time points. The origin of the stability classification system is clearly visible in these plots, with a survival fraction of 0.5 marking the boundary between two stability classes in each plot. 

\begin{figure*}
	\centering
    \includegraphics[width=\textwidth]{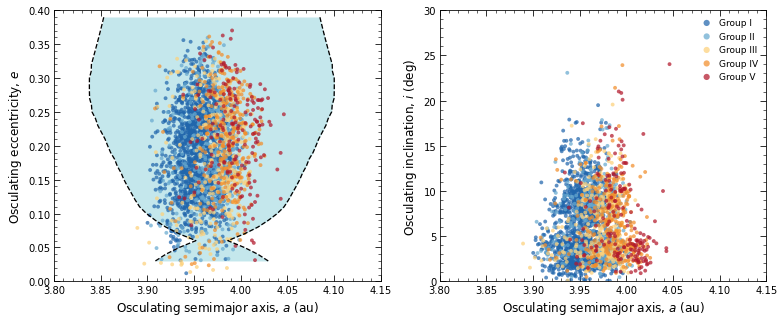}
  \caption{The distribution of the Hildas studied in this work, reprising the information presented in the left-hand panels of Figure 1, but with the individual Hildas now coloured to indicate in which stability classes they fall. The left-hand panel shows the initial semi-major axes and eccentricities for each of the 2,699 Hildas studied in this work, along with the structure of the Jovian 3:2 mean-motion resonance (the shaded region), calculated for an orbital inclination of 5.5$^\circ$, the mean inclination of the orbits of the Hildas studied, using the methodology outlined in \citet{Gallardo2020}. The right-hand panel shows the same objects plotted in semi-major axes vs orbital inclination space. Interestingly, a pronounced trend is visible where the most unstable Hildas studied in this work are typically found at larger initial semi-major axes.}
  \label{osculating_vs_groups}
\end{figure*}


\begin{figure*}
	\centering
    \includegraphics[width=\textwidth]{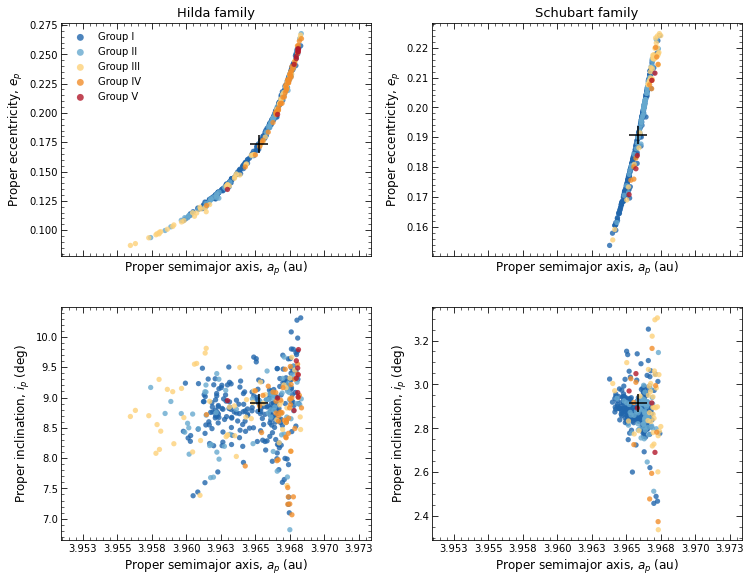}
  \caption{The initial distribution of all identified members of the Hilda and Schubart collisional families that were studied in this work, in proper element space. The upper panels show the Hilda family (left) and Schubart family (right) in proper semi-major axis vs eccentricity space, whilst the lower panels plot proper semi-major axis against proper inclination. The locations of (153) Hilda (left panels) and (1911) Schubart (right panels) is marked by the + symbol. It should be noted that the Schubart family occupies a smaller region in both plots, which suggests a younger age (as the members have had less time to disperse). The colours of the individual points denote the stability classes for each member of the family, based on our simulations. For both families, it appears that most (but not all) of the unstable members (Groups IV and V) are found towards the edges of the family grouping, a finding that is most apparent in the lower panels.}
  \label{fams_zoom}
\end{figure*}


Figure~\ref{osculating_vs_groups} shows the initial osculating orbital elements for all Hildas studied in this work, reprising the information shown in Figure~\ref{fig:orbital_elements}, but with each individual Hilda coloured according to its dynamical stability classification obtained from our simulations. It is interesting to note that the less stable Hildas (groups III, IV, and V) are located, in the main, in the outer half of the sample studied (i.e. exterior to $\sim$ 3.97 au), whilst the stable objects are typically found in the interior. To demonstrate this, in Table~\ref{tab:stabpercentages}, we detail the number of members of each group that began our simulations interior to and exterior to 3.97 au, along with the fraction of the population in each group that began in those locations. Almost 80\% of the members of Group I (the Strongly Stable Hildas) began our simulations with orbital semi-major axes less than or equal to 3.97, whilst only 15\% of Group V and 22.3\% of Group IV objects began the simulations interior to that line. 

\begin{table}
\centering
\caption{Table detailing the number of Hildas in each stability group (from the most stable, Group I, to the least, Group V, as per Table~\ref{tabla:classif_scheme}) that had initial orbital semi-major axes interior to 3.97 au (N$_{\textrm{int}}$), or exterior to that location (N$_{\textrm{ext}}$). The final two columns detailed the fraction of each group found interior to ($f_{\textrm{int}}$) or exterior to ($f_{\textrm{ext}}$) that line. It is immediately apparent that the least stable members of the population are preferentially found at higher semi-major axes in our sample.}
\label{tab:stabpercentages}
\begin{tabular}{cccccc}
\hline
Group & N$_{\textrm{tot}}$ & N$_{\textrm{int}}$ & N$_{\textrm{ext}}$ & $f_{\textrm{int}}$ & $f_{\textrm{ext}}$ \\
\hline
I & 1475 & 1177 & 298 & 0.798 & 0.202 \\
II & 408 & 179 & 229 & 0.439 & 0.561 \\
III & 327 & 129 & 198 & 0.395 & 0.605 \\
IV & 309 & 69 & 240 & 0.223 & 0.777 \\
V & 180 & 27 & 153 & 0.150 & 0.850 \\
\hline
\end{tabular}
\end{table}

Figure~\ref{fams_zoom} shows the currently known members of the Hilda (left) and Schubart (right) collisional families, distributed in proper element space, with the individual members again coloured based on the dynamical stability group into which they fall. We note in passing that the spread of the Schubart family is significantly smaller than that of the Hilda family, which is in keeping with the findings of \citet{Broz-Hildas-LHB}, who suggest that the Hilda collisional family is much older than that belonging to (1911) Schubart. For both families, the majority of unstable members appear to be located to the periphery of the region occupied by the bulk of the family members -- something which is most obvious in the lower two panels, which plot the proper semi-major axes and inclinations of the family members.

\subsection{Dynamical Half-lives}
\label{subsec:half-life}

\begin{figure}
	\includegraphics[width=\columnwidth]{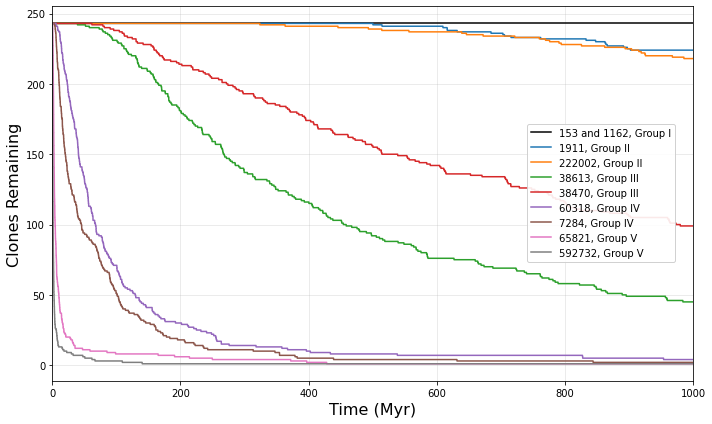}
  \caption{The number of surviving clones versus time for two representative asteroid examples from each stability group as described in Table~\ref{tabla:classif_scheme}. In addition to the clear differences in decay rates between members of the different groups, it is immediately apparent that the least stable objects (from groups IV and V) exhibit immediate decay, whilst the members of the stable groups (I -- III) show a pronounced delayed start to their decay.}
  \label{3_examples_group}
\end{figure}

Figure~\ref{3_examples_group} shows the rate of decay of the 243 clones for two representative asteroids from each stability group, highlighting the typical dynamical evolution observed for objects across our classification scheme. It is immediately apparent that the behaviour of the examples of the unstable objects (from Groups IV and V) is markedly different to those from the stable sample (Groups I -- III). In addition to the more rapid decay exhibited by the unstable objects, it is clear that their decay starts immediately the simulations begin, with no appreciable delay. By contrast, the objects in the stable populations exhibit a delayed start to their decay\footnote{Plots showing the decay in population for all objects in stability groups III, IV and V can be found in Appendix~\ref{sec:appendix_plots}\referee{, whilst Appendix~\ref{sec:appendix_elemperturb} shows that the stability (or lack of stability) of the objects in our sample was independent of the direction in which the individual orbital elements were perturbed}.}.

\begin{table}
\centering
\caption{Summary of the dynamical properties of selected Hilda asteroids. The table presents the asteroid number, the number of surviving clones after the integration period, and the estimated half-life in gigayears (Gyr). The values provide insight into the long-term stability and dynamical characteristics of these objects.}
\label{big_table}
\begin{tabular}{ccc}
\hline
\textbf{Asteroid} & \textbf{Surviving Clones} & \textbf{Half-life (Gyr)} \\
\hline
(153) Hilda  & 243 & $\infty$ \\
(190) Ismene  & 170 & 1.94 \\
(361) Bononia  & 243 & $\infty$ \\
(499) Venusia  & 0   & 0.12 \\
(748) Sime{\"i}sa  & 243 & $\infty$ \\
(958) Asplinda  & 99  & 0.77 \\
(1038) Tuckia & 242 & 169\\
(1162) Larissa & 243 & $\infty$ \\
(1202) Marina & 243 & $\infty$ \\
(1212) Francette & 243 & $\infty$ \\
(1268) Libya & 243 & $\infty$ \\
(1269) Rollandia & 243 & $\infty$ \\
(1345) Potomac & 11  & 0.22 \\
(1529) Oterma & 0                 & 0.126 \\
(1911) Schubart & 224 & 8.51 \\
\hline
\end{tabular}
\end{table}

Previous studies of Solar system small bodies have used the concept of a `dynamical half-life' to compare the stability of different objects \citep[e.g.][]{Dones-1996-half-lifes,Horner-2004,Horner-2004b,LC18,Holt20,Wood-2023-Uranus_Trojans}. The use of such a construct is driven by the similarity between the decay behaviour shown in Figure~\ref{3_examples_group} and the decay of radioactive materials over time. Here, we follow \citet{Horner-2004}, and define the dynamical half-life, $T_{1/2}$, as follows:

\begin{equation}
N = N_0 e^{-\lambda t} \quad \text{and} \quad \lambda = \frac{0.693}{T_{1/2}}
\end{equation}

Where $\lambda$ is the decay constant, $T_{1/2}$ the half-life, $N_0$ the initial number of particles, and $N$ the number of surviving test particles after elapsed time $t$.

Although exponential decay models are typically associated with unstable or chaotic systems, their application to Hilda asteroids is supported by previous research. \citet{Franklin-1993-Chaotic_Orbits} demonstrated that many Hildas exhibit short Lyapunov times ($T_L \sim 10^4$--$10^5$ yr), indicative of local chaos, yet can maintain orbital coherence over gigayear timescales. This seemingly paradoxical behaviour is explained through the empirical correlation between Lyapunov time and dynamical lifetime proposed by \citet{lecar1992}, whereby even chaotic orbits may persist for billions of years. These findings provide a robust theoretical basis for interpreting clone loss curves through half-life modelling, even in regions where dynamical chaos is present.

\begin{figure}
	\centering
    \includegraphics[width=\columnwidth]{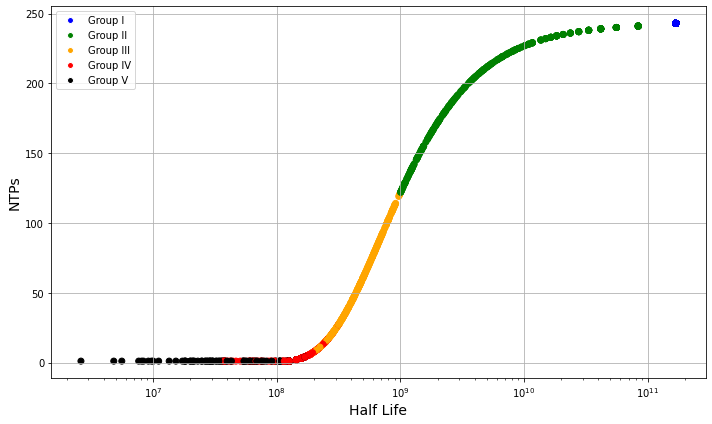}
  \caption{The relationship between orbital stability expressed as half-life ($T_{1/2}$) and the number of surviving test particles (NTPs) as surviving clones. The stability groups are clearly distinguished: Group I (blue) corresponds to highly stable orbits (243 clones), Group II (green) to moderately stable orbits (122–242 clones), Group III (orange), which notably appears centred at approximately 1 Gyr, indicates intermediate orbital stability, and Groups IV (red) and V (black) represent the least stable orbits. This visualisation highlights the continuous stability transition within the studied particle population.}
  \label{Half_life}
\end{figure}

Figure~\ref{Half_life} presents the relationship between the dynamical half-life and the number of clones that survived through to the end of the simulations at the 1 Gyr mark. As a visual aid to the reader, we have colour-coded the data, showing where the five dynamical classes presented in Table~\ref{tabla:classif_scheme} sit in the plot. It is immediately clear that there is no clear discontinuity -- revealing that the choice of boundaries between the five classes is somewhat arbitrary, and that there is no clear and sharp dividing line between the stable and unstable members of the Hilda population that speaks to two (or more) distinct groups of objects in one shared space.

Previous studies have shown that the dynamical structure of the Jovian 3:2 mean-motion resonance is also influenced by secular and secondary resonances. In particular, the high-eccentricity boundary of the stable region is affected mainly by the $\nu_5$ and $\nu_6$ secular resonances, whereas secondary resonances become important at low eccentricities \citep{Ferraz-Mello1988,Morbidelli-1993_sec_resonan_MMR,Broz-Hildas-LHB}. The gravitational effects associated with these resonances are naturally included in our full $N$-body integrations. However, identifying the contribution of individual secular or secondary resonances would require a dedicated frequency analysis, which will be addressed in the next paper of this series (Paper II).

\section{Discussion and Conclusions}
\label{Sec:Conclusions}

We performed detailed 1~Gyr N-body simulations of the numbered Hilda asteroid population under the gravitational influence of the Sun and the four giant planets (Jupiter, Saturn, Uranus, and Neptune). For each of the 2,699 selected asteroids, we generated test particles by varying five orbital elements ($a$, $e$, $i$, $\omega$, and $\Omega$) within their one-sigma uncertainties, resulting in a total of 655,857 clones (243 clones per Hilda studied). 

These long-term forward integrations provide a robust timescale to assess the dynamical half-lives and survival rates of the clones, enabling an evaluation of the long-term orbital stability of each asteroid. This, in turn, allows us to investigate whether the Hilda population as a whole displays behaviour that is consistent with a primordial origin -- in other words, whether its members have resided in the 3:2 mean-motion resonance with Jupiter since the formation of the Solar system -- or whether some members were captured more recently, and are transient interlopers to the population.

We find that more than 54\% of the Hilda asteroids studied in this work move on orbits that are dynamically stable on Gyr timescales, remaining confined within the 3:2 resonance throughout the integration. A substantial fraction also display quasi-stable behaviour, consistent with weakly chaotic motion or slow diffusion, in agreement with prior studies. In numerical terms, 1,475 of the 2,699 Hildas examined in this work were found to be 'Strongly Stable Hildas', with all 243 of their clones surviving until the end of our simulations. A further 408 Hildas saw more than 50\% of their clones survive for the full duration of our simulations (being classified herein as 'Stable Hildas'). These two stability classes combined account for 69.75\% of the total sample of Hildas studied. Such a high fraction reinforces the established wisdom that the Hilda population constitutes a highly stable dynamical reservoir. Furthermore, the predominance of extremely long-lived objects is consistent with the idea that a substantial fraction of the population has remained within Jupiter's 3:2 mean-motion resonance since Solar system's birth, or at least since the large-scale migration of the giant planets came to an end.

It is worth noting that the least stable objects (those in groups IV and V) begin to decay from the very start of our simulations, whilst those that are more stable, but still see ejections (groups II and III), feature a delayed start to their decay. Such behaviour has been described in previous dynamical studies of moderately unstable members of resonant populations \citep[e.g.][]{QR322,LC18,Anchises}, and this delayed initiation of decay seems to be the hallmark of objects that are likely to be primordial members of the population. Those objects whose decay exhibits no delay are almost certainly true interlopers to the domain of the Hildas, and it will be interesting to see whether future observational studies reveal any clear difference in properties between these classes of object.

Overall, our results align with earlier findings by \citet{Franklin-1993-Chaotic_Orbits}, who studied several of the more famous members of the Hilda population known at the time. They demonstrated that those Hildas, whilst exhibiting some degree of chaotic behaviour, had orbital stabilities that were consistent with their survival in the Hilda population since the Solar system's youth. 

Additional support comes from the work of \citet{Dahlgren1993}, who found no evidence of chaos in his sample and suggested that Hildas can survive in stable orbits for over 1~Myr. In a follow-up, \citet{Dahlgren-1998__Hildas_vel} showed that Hildas (with $D \geq 50$~km) have the lowest mean collision probabilities among the major asteroid groups, indicating reduced collisional evolution. This dynamical and collisional quiescence aligns with the quasi-integrable structure found in recent theoretical work by \citet{Asano-2024}, who modelled the Sun--Jupiter--Hilda system using a single-resonance Hamiltonian, confirming the long-term regularity of 3:2 resonant dynamics. Our findings are also corroborated by \citet{Zain-DiSisto-2025}, whose collisional and dynamical simulations show that Hildas experience significantly fewer disruptions than quasi-Hildas (their terminology for objects temporarily captured to Hilda-like orbits), primarily because the latter undergo more frequent encounters with Jupiter. This underscores the Hildas' greater dynamical insulation and supports our observed trends in high survival fractions and longer half-lives.

Our finding that more than half of the numbered Hildas show extremely long-term dynamical stability is well aligned with the recent population-level study of \cite{Vokrouhlicky-2025-Hildas-mag-dist}, who identify a well-defined, low-inclination and low-eccentricity core within the Hilda population, whose orbital and magnitude distribution suggest preservation over gigayear timescales. Their statistical inference of a long-lived primordial core is fully consistent with our direct N-body integrations, which show that these objects occupy the deepest and most stable regions of the 3:2 resonance. The convergence of these independent studies strengthens the idea that the Hilda population has a robust stable core surrounded by a dynamically hotter outer component.

This interpretation is strengthened by comparisons with other resonances. \citet{Morbidelli-1993_sec_resonan_MMR} showed that the Hilda region is protected from secular resonances like $\nu_5$ and $\nu_6$ by phase-locking mechanisms, while \citet{Nesvorny-1997_ast_pop_1ordJUP} found that chaotic diffusion in the 3:2 resonance is two orders of magnitude slower than in the 2:1. \citet{Franklin-Hildas-capturados2004} noted that Hilda-like orbits remain stable for up to 4.5~Gyr, even when mildly chaotic. Meanwhile, \citet{Gaspar} and \citet{SlizBalogh-2023_Jup_Organizer} demonstrated that a fraction of primordial bodies can be retained in the 3:2 resonance during a jumping-Jupiter scenario, reinforcing the plausibility of early capture followed by long-term stability.

Whilst the majority of Hildas examined in this study exhibit strong dynamical stability on Gyr timescales, we find that a surprising number of objects in our sample escape from the Hilda population on timescales of just a few million years. Indeed, 180 of the Hildas studied in this work saw more than 50\% of all clones removed from the Solar system entirely within the first 10 Myr of the simulations. We consider such objects to be 'interlopers' -- they are most likely objects temporarily captured to the Hilda population in the relatively recent past. Upon escaping the 3:2 mean-motion resonance, those objects may join either the short-period comet or near-Earth asteroid populations, differentiated by whether they exhibit outgassing and cometary activity or remain inert. 

Regarding the origin of the interlopers, it seems reasonable to speculate that such temporary captures would be sourced from both the short-period comet and near-Earth asteroid populations. Indeed, \cite{Cheng_qHilda2013} state that quasi-Hilda comets provide a direct analogue for such transport, finding that 212P/2000 YN$_{30}$ could have evolved from a Centaur-like orbit into its current quasi-Hilda configuration. Moreover, \citet{Correa-Otto2024} identified 47 quasi-Hilda candidates whose backward orbital evolution is consistent with recent injection from the Centaur population. Such an origin would mirror the idea that Centaurs can be captured to both the Jovian and Neptunian Trojan populations \citep[e.g.][]{Karl04,centaur_trojan,KV18,Guan12,Greenstreet2020,Green24}. Indeed, there are a number of short-period comets (such as the aforementioned 212P) that currently move on orbits whose elements place them firmly amongst the Hilda population, with orbital periods close to 3:2 commensurability with Jupiter (though it should be noted that simply having an orbital period that is close to 3:2 commensurability is not the same as being trapped, temporarily or otherwise, in the 3:2 mean-motion resonance). It would be interesting to see whether, in the years to come, any of the 180 interlopers we identify in this work are found to exhibit low levels of cometary activity. \referee{In particular, the interlopers (452814) 2006 PE$_{23}$, (450801) 2007 TP$_{383}$, and (162232) 1999 TC$_{154}$, would be potential targets for such observations, having the lowest current perihelion distances amongst the interloper objects, at 2.514~au, 2.537~au, and 2.562~au, respectively.}

The significant number of Hildas that show some attrition through the 1 Gyr simulations, but with dynamical half-lives in excess of 100 Myr, are an indication that the primordial Hilda population was almost certainly larger than that we see today. This interpretation is consistent with the work of \citet{wong17}, who suggest that the shallower absolute magnitude distribution observed for the Hildas when compared to the Jovian Trojans implies that an initially larger Hilda population was depleted as collisions preferentially drove smaller bodies out of the comparatively narrow 3:2 resonance. Whilst our simulations do not include collisional evolution or size-dependent forces, they independently demonstrate the existence of a long-term dynamical leakage mechanism capable of contributing to such depletion.

As has previously been discussed for both the Jovian and Neptunian populations \citep[e.g.][]{QR322,Anchises}, the presence of objects whose dynamical lifetimes are measured in hundreds of millions of years is entirely compatible with such objects being relics of a once larger population. As time passes, such objects gradually diffuse out of the Hilda population, escaping from the 3:2 mean-motion resonance with Jupiter, and moving onto dynamically unstable orbits reminiscent of near-Earth asteroids and short-period comets. Our results therefore suggest that the primordial Hilda population has served as a source of material to the Solar system's unstable small body populations - joining the other resonant populations as potential contributors to the impact flux at Earth \citep[e.g.][]{DiS2005,JH_SPC,HL2010}.

\begin{table}
\centering
\caption{Median proper orbital elements for the five stability groups,
for the overall proper-element sample and for the Hilda and Schubart
collisional families.}
\label{tab:group_medians}
\begin{tabular}{lrrrr}
\hline
Group & Number & $a_{\rm med}$ (au) & $e_{\rm med}$ & $i_{\rm med}$ (deg) \\
\hline

\multicolumn{5}{l}{\textit{Overall sample}} \\
I & 1218 & 3.965515 & 0.180664 & 4.042 \\
II & 325 & 3.965926 & 0.188391 & 4.800 \\
III & 256 & 3.966145 & 0.191276 & 5.119 \\
IV & 198 & 3.967204 & 0.218629 & 7.361 \\
V & 76 & 3.967814 & 0.239650 & 8.868 \\
\hline
\multicolumn{5}{l}{\textit{Hilda family}} \\
I & 230 & 3.965495 & 0.177374 & 8.820 \\
II & 67 & 3.966040 & 0.188002 & 8.845 \\
III & 55 & 3.963626 & 0.147051 & 8.788 \\
IV & 47 & 3.967233 & 0.221606 & 8.750 \\
V & 10 & 3.968021 & 0.250134 & 9.200 \\
\hline
\multicolumn{5}{l}{\textit{Schubart family}} \\
I & 237 & 3.965783 & 0.187936 & 2.880 \\
II & 63 & 3.966281 & 0.200110 & 2.878 \\
III & 37 & 3.966928 & 0.216194 & 2.896 \\
IV & 10 & 3.966760 & 0.206935 & 2.850 \\
V & 5 & 3.965794 & 0.183883 & 2.916 \\

\hline
\end{tabular}
\end{table}

Whilst our simulations do not investigate how the Hildas were originally trapped in the 3:2 mean-motion resonance, so do not differentiate between primordial objects and those captured during the migration of the giant planets, they do demonstrate that the majority of the known Hilda population exhibit dynamical stability that is entirely compatible with their having been trapped in the 3:2 mean-motion resonance with Jupiter since the formation of the planets. The number of interloper objects found among the numbered Hildas is a fascinating outcome of our work, and we look forward to future observational studies that will allow for the comparison between the interlopers and strongly stable Hildas, to look for evidence of physical similarities and differences between the two populations. In particular, it will be interesting to see whether such studies reveal any evidence of cometary activity among the more unstable Hildas -- something that would identify objects captured from the short-period comet population, with their origins at far greater heliocentric distance than the outer reaches of the asteroid belt.


\section*{Acknowledgements}

This work has made use of the NASA/JPL Solar system Dynamics page, at \url{https://ssd.jpl.nasa.gov/}. It has also made use of the AstDyS database, at \url{https://newton.spacedys.com/astdys/index.php?pc=3.0}, and NASA’s Astrophysics Data System, ADS, at \url{https://ui.adsabs.harvard.edu/}.

The lead author wishes to thank Mr. Rafael Fuentes, software engineer, for his valuable advice and insightful discussions on topics related to computational science.

\referee{The authors would like to express their sincere gratitude for the exceptionally helpful,  detailed, and swift feedback from the referee, Matthew Belyakov, which helped us to significantly improve this work from our first submission to the final published paper. }

\section*{Data Availability}

The complete machine-readable catalogue containing all 2,699 numbered Hilda asteroids analysed in this work is publicly available through the project GitHub repository.\footnote{\url{https://github.com/cchavez-astro/Hildas_table_cchavez}}



\bibliographystyle{mnras}
\bibliography{references} 



\appendix
\newpage
\section{Tables by Stability Group}
\label{sec:appendix_tables}

This Appendix lists the first ten asteroids of each stability group described in Section~\ref{Sec:Stability}.

\begin{table*}
\centering
\small
\caption{Group I: The first 10 `Strongly Stable' Hilda asteroids in our simulations, sorted by the number given to the asteroids by the IAU. The columns give the asteroid number (ID), the number of test particles remaining after 1 Gyr (NTP), and the dynamical half-life ($T_{1/2}$). See Section~\ref{Sec:Stability}.}
\label{tab:G1_first10}
\begin{tabular}{crr}
\hline
\textbf{ID} & \textbf{NTPs} & \boldmath{$T_{1/2}$} \textbf{(yr)}\\
\hline
(153) Hilda & 243 & $\infty$ \\
(361) Bononia & 243 & $\infty$ \\
(748) Sime{\"i}sa & 243 & $\infty$ \\
(1162) Larissa & 243 & $\infty$ \\
(1202) Marina & 243 & $\infty$ \\
(1212) Francette & 243 & $\infty$ \\
(1268) Libya & 243 & $\infty$ \\
(1269) Rollandia & 243 & $\infty$ \\
(1512) Oulu & 243 & $\infty$ \\
(1578) Kirkwood & 243 & $\infty$ \\
\hline
\end{tabular}
\end{table*}

\begin{table*}
\centering
\small
\caption{Group II: the first 10 `Stable' Hilda asteroids in our sample, with the number of surviving clones at 1 Gyr and the dynamical half-life. See Section~\ref{Sec:Stability}.}
\label{tab:G2_first10}
\begin{tabular}{crr}
\hline
\textbf{ID} & \textbf{NTPs} & \boldmath{$T_{1/2}$} \textbf{(yr)}\\
\hline
(190) Ismene & 170 & 1.940e+09 \\
(1038) Tuckia & 242 & 1.690e+11 \\
(1180) Rita & 137 & 1.209e+09 \\
(1911) Schubart & 224 & 8.512e+09 \\
(3134) Kostinsky & 139 & 1.241e+09 \\
(6124) Mecklenburg & 167 & 1.848e+09 \\
(6984) Lewiscarroll & 241 & 8.385e+10 \\
(8376) & 234 & 1.836e+10 \\
(10331) Peterbluhm & 141 & 1.273e+09 \\
(10889) & 208 & 4.456e+09 \\
\hline
\end{tabular}
\end{table*}

\begin{table*}
\centering
\small
\caption{Group III: first 10 Hilda asteroids identified as `Moderately Stable Hildas' in this work, with the number of surviving clones at 1 Gyr and the dynamical half-life. See Section~\ref{Sec:Stability}.}
\label{tab:G3_first10}
\begin{tabular}{crr}
\hline
\textbf{ID} & \textbf{NTPs} & \boldmath{$T_{1/2}$} \textbf{(yr)}\\
\hline
(958) Asplinda & 99 & 7.718e+08 \\
(1439) Vogtia & 60 & 5.134e+08 \\
(1941) Wild & 32 & 3.471e+08 \\
(8551) Daitarabochi & 79 & 6.168e+08 \\
(11750) & 120 & 9.939e+08 \\
(12006) Hruschka & 3 & 1.687e+08 \\
(12920) & 9 & 2.103e+08 \\
(13381) & 64 & 5.194e+08 \\
(13897) Vesuvius & 77 & 6.098e+08 \\
(15638) & 84 & 6.524e+08 \\
\hline
\end{tabular}
\end{table*}

\begin{table*}
\centering
\small
\caption{Group IV: first 10 Hildas identified as `Unstable Hildas' in this work. The columns show the number of surviving clones after 1 Gyr and the dynamical half-life. See Section~\ref{Sec:Stability}.}
\label{tab:G4_first10}
\begin{tabular}{crr}
\hline
\textbf{ID} & \textbf{NTPs} & \boldmath{$T_{1/2}$} \textbf{(yr)}\\
\hline
(1345) Potomac & 11 & 2.239e+08 \\
(1529) Oterma & 0 & 1.262e+08 \\
(2483) Guinevere & 20 & 2.775e+08 \\
(2760) Kacha & 3 & 1.687e+08 \\
(3557) Sokolsky & 5 & 1.872e+08 \\
(3655) Eupraksia & 2 & 1.687e+08 \\
(3843) OISCA & 5 & 1.784e+08 \\
(4317) Garibaldi & 2 & 1.444e+08 \\
(7174) Semois & 1 & 1.577e+08 \\
(7284) & 2 & 1.444e+08 \\
\hline
\end{tabular}
\end{table*}

\begin{table*}
\centering
\small
\caption{Group V: first 10 members of the `Interloper' class amongst the Hilda population, as identified in this work. The columns show the number of surviving clones at 1 Gyr and the dynamical half-life. See Section~\ref{Sec:Stability}.}
\label{tab:G5_first10}
\begin{tabular}{crr}
\hline
\textbf{ID} & \textbf{NTPs} & \boldmath{$T_{1/2}$} \textbf{(yr)}\\
\hline
(499) Venusia & 0 & 1.187e+08 \\
(4446) Carolyn & 0 & 1.262e+08 \\
(11274) Castillo-Rogez & 0 & 9.172e+06 \\
(13504) & 1 & 1.444e+08 \\
(17305) Caniff & 0 & 4.256e+07 \\
(22058) & 1 & 1.262e+08 \\
(23186) Knauss & 0 & 6.651e+07 \\
(40238) & 0 & 2.744e+07 \\
(45739) & 0 & 1.577e+08 \\
(54514) & 0 & 1.262e+08 \\
\hline
\end{tabular}
\end{table*}

\newpage
\section{Decay curves for all Hildas in Groups III, IV, and V}
\label{sec:appendix_plots}

This Appendix presents plots showing the decay in the number of surviving test particles across the first 200 Myr of our simulations for all objects that fall in stability groups III, IV, and V. 

\begin{figure}
\centering
\includegraphics[width=\columnwidth]{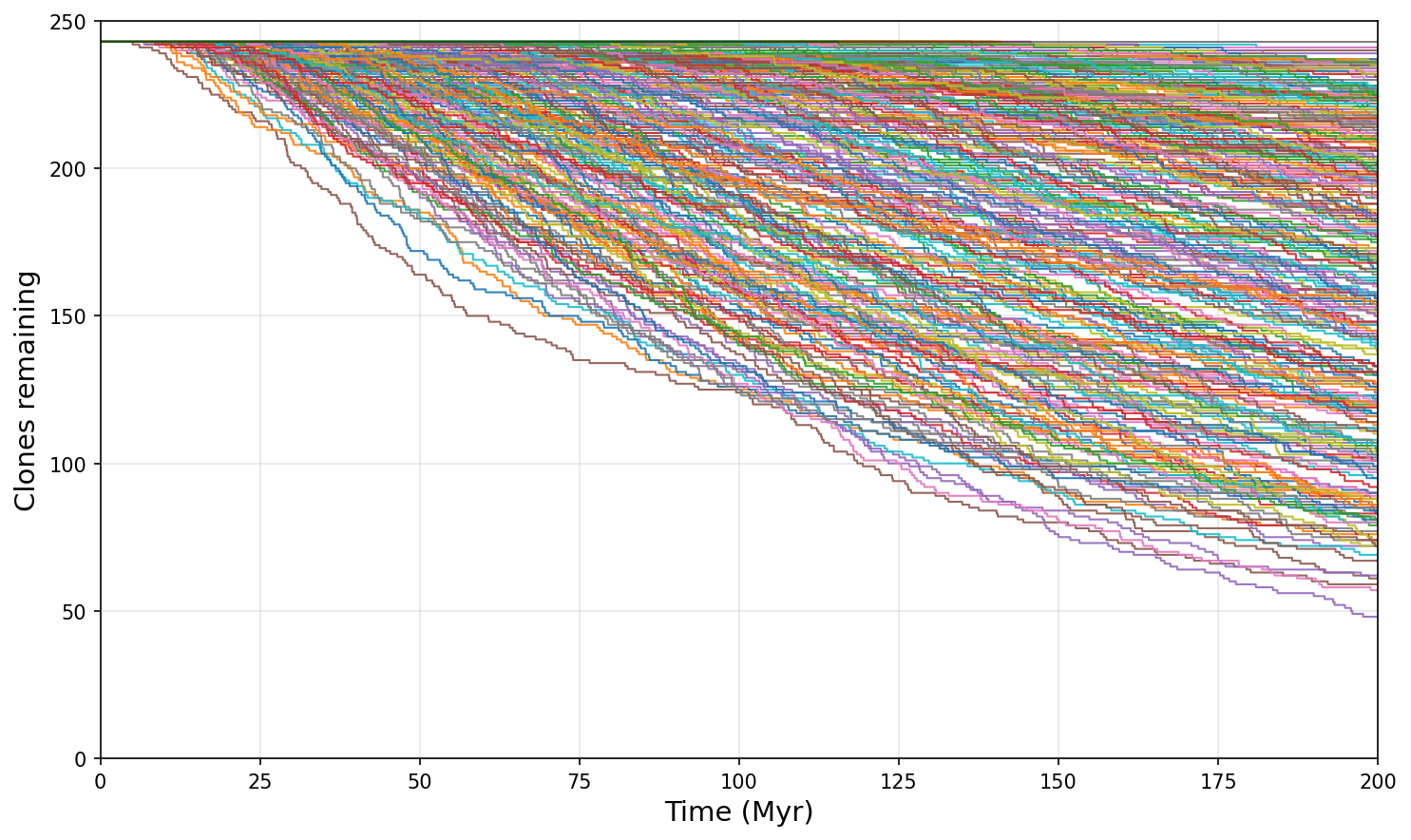}\\
(a) Group III (Moderately Stable Hildas)
\medskip
\includegraphics[width=\columnwidth]{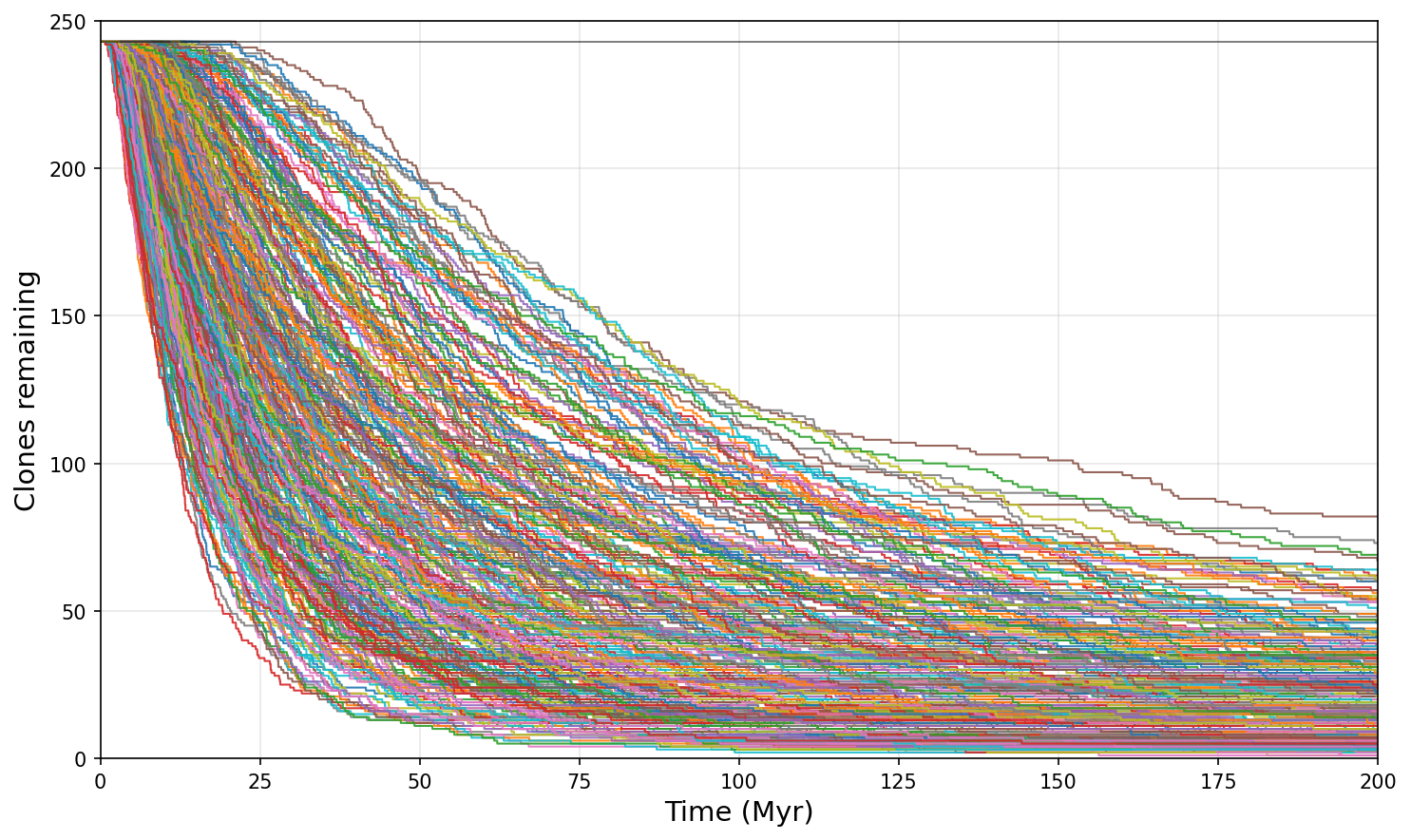}\\
(b) Group IV (Unstable Hildas)
\medskip
\includegraphics[width=\columnwidth]{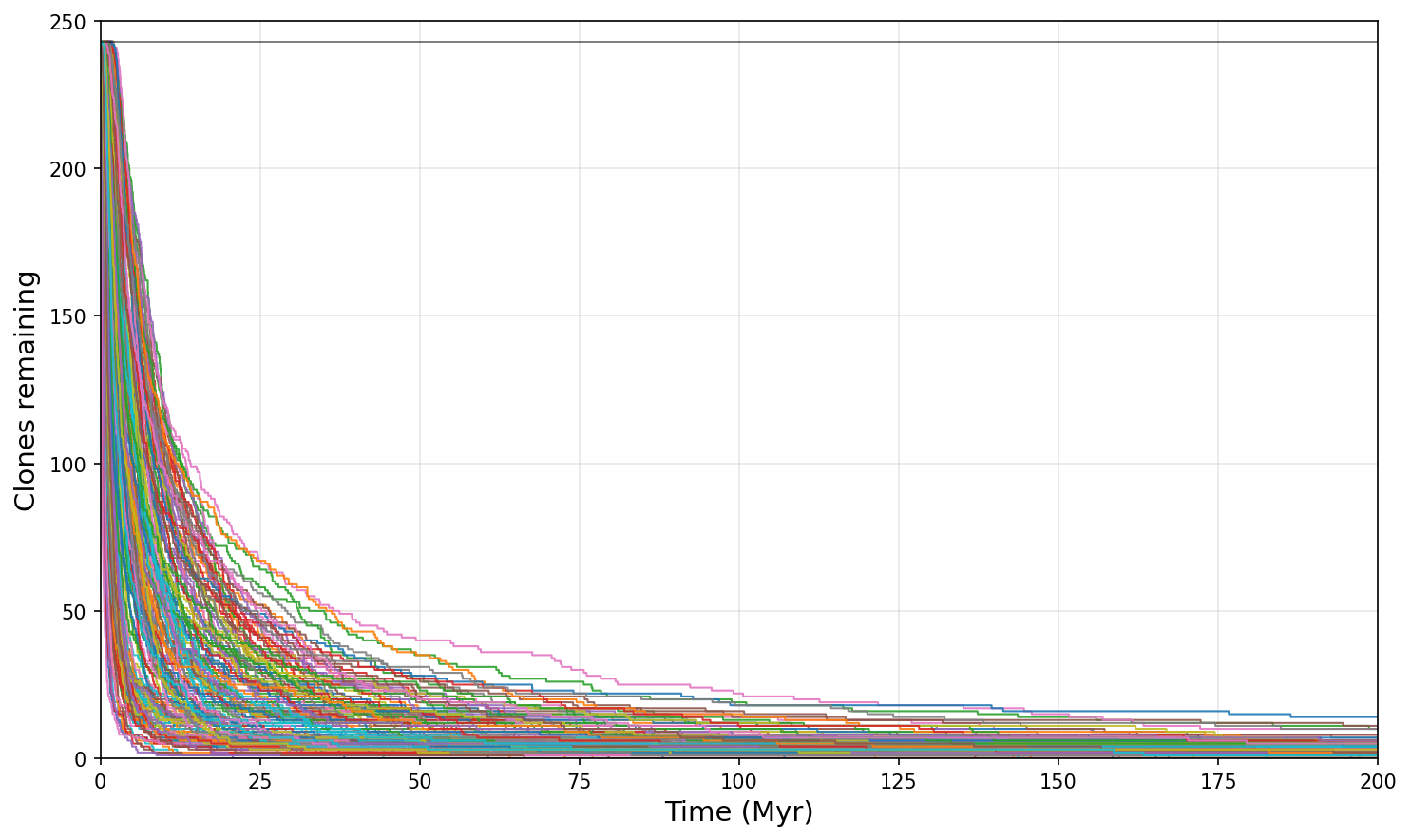}\\
(c) Group V (Interlopers)
\caption{Early-stage clone-survival (0–200 Myr) for Hildas in the unstable classes of groups III, IV and V, showing the number of surviving test particles per asteroid versus time. Each coloured curve corresponds to one asteroid; the horizontal line marks the initial 243 clones. Panels: (a) Group III — A slower decay, with the majority of clones still present at 100 Myr and only moderate attrition by 200 Myr; (b) Group IV — A certain rapid early depletion, with most asteroids dropping below $\sim$50\% well before 100 Myr; (c) Group V — survival typically collapses within $\lesssim$10 Myr, indicating very short residence in the 3:2 resonance. The sequence illustrates the monotonic progression in stability encoded in our classification scheme (see Table~\ref{tabla:classif_scheme}).}
\label{fig:unstable-groups}
\end{figure}

\newpage
\section{Survival Fractions as a Function of Perturbed Orbital Element}
\label{sec:appendix_elemperturb}

\referee{This Appendix compares the number of lost and surviving clones as a function of the orbital elements altered in the cloning process. In total, 655,857 individual Hilda clones were simulated in our work (2,699 Hildas, each cloned a total of 243 times, with three clones in each of the five orbital elements $a-e-i-\Omega-\omega$. In the table below, we detail the total number of clones for each increment of each element (e.g. for $a - 1\sigma$, $a$, and $a + 1\sigma$; with 218,619 clones each), then the number of those clones removed from the simulations (through collision or ejection), the number that survive, and the surviving fraction. We note that, for each element studied, there was no statistically significant difference in the survival fraction across the three increments for that element. This is not unexpected -- all Hildas studied in this work are numbered asteroids, with orbits that are known to very high precision. As a result, the scale of increments in each of the orbital elements used is very small.}

\begin{table*}
\centering
\small
\caption{The number of Hilda clones created for each particular increment of each orbital element (N$_{init}$), along with the number of those clones that were lost during our simulations (through ejection or collision; N$_{lost}$), and the number and fraction that survived (N$_{surv}$; $f_{surv}$).}
\label{tab:elemperturb}
\begin{tabular}{cccccc}
\hline
\textbf{Element} & \textbf{Perturbation} & \textbf{N$_{init}$} & \textbf{N$_{lost}$} & \textbf{N$_{surv}$} & \textbf{$f_{surv}$}\\
\hline
$a$ & $-1\sigma$ & 218,619 & 65,454 & 153,165 & 0.700602 \\
$a$ & 0          & 218,619 & 65,535 & 153,084 & 0.700232 \\
$a$ & $+1\sigma$ & 218,619 & 65,536 & 153,083 & 0.700227 \\
\hline
$e$ & $-1\sigma$ & 218,619 & 65,499 & 153,120 & 0.700397 \\
$e$ & 0          & 218,619 & 65,569 & 153,050 & 0.700076 \\
$e$ & $+1\sigma$ & 218,619 & 65,457 & 153,162 & 0.700589 \\
\hline
$i$ & $-1\sigma$ & 218,619 & 65,512 & 153,107 & 0.700337 \\
$i$ & 0          & 218,619 & 65,435 & 153,184 & 0.700689 \\
$i$ & $+1\sigma$ & 218,619 & 65,578 & 153,041 & 0.700035 \\
\hline
$\omega$ & $-1\sigma$ & 218,619 & 65,568 & 153,051 & 0.700081 \\
$\omega$ & 0          & 218,619 & 65,441 & 153,178 & 0.700662 \\
$\omega$ & $+1\sigma$ & 218,619 & 65,516 & 153,103 & 0.700319 \\
\hline
$\Omega$ & $-1\sigma$ & 218,619 & 65,474 & 153,145 & 0.700511 \\
$\Omega$ & 0          & 218,619 & 65,588 & 153,031 & 0.699989 \\
$\Omega$ & $+1\sigma$ & 218,619 & 65,463 & 153,156 & 0.700561 \\
\hline
\end{tabular}
\end{table*}

\bsp	
\label{lastpage}
\end{document}